\documentstyle[aps]{revtex}
\begin{document}
\title{Realistic model of correlated disorder and Anderson localization}

\author{V.V. Flambaum}
\address{ School of Physics, University of New South Wales,
Sydney 2052, Australia}
\maketitle

\date{\today}
\begin{abstract}
 A conducting 1D line or 2D plane inside (or on the surface of) an insulator
is considered. Impurities displace the charges inside the
insulator. This  results in a long-range fluctuating electric field
acting on the conducting line (plane). This field can be modeled
by that of randomly distributed electric dipoles. This model
provides a random correlated potential with $\langle U(r)U(r+k)\rangle 
\propto 1/k$.
In the 1D case such correlations give essential corrections to the localization
length but do not destroy Anderson localization.


\end{abstract}



It was recently stated in \cite{Izrailev1998} that some special correlations
in a random potential can produce a mobility edge (between localized
and delocalized states)
inside the allowed band in the 1D tight-binding model. In principle,
extrapolation of this result to 2D systems may give a possible
explanation of the insulator-conductor transition in dilute
2D electron systems observed in ref. \cite{Kravchenko1996}.
In such a situation it is very important to build a reasonable
model of ``correlated disorder'' in real systems and calculate
the effects of this ``real'' disorder.

    Usually, a 1D or 2D conductor is made inside or on the surface
of an insulating material. Impurities inside the insulator 
displace the electric charges. However, a naive ``random charge''
model violates electro-neutrality and gives wrong results. 
Indeed, the impurities do not produce new charges, they only
displace charges thus forming electric dipoles. Therefore,
we consider a model of randomly distributed electric dipoles
(alternatively, one can consider a spin glass model which
gives the correlated random magnetic field). The dipoles have long-range
electric field. Therefore, the potentials at different sites
turn out to be correlated.

  The potential energy produced by the system of the dipoles $d_j$ is equal to 
\begin{equation}
\label{U}
U(r)= e\sum_j {\bf d_j \nabla} \frac{1}{|\bf{r-R_j}|} .
\end{equation}
The average value of this potential is zero if $\langle \bf {d_j} \rangle =0$.
 The fluctuations of the potential at a given site are as follows
\begin{equation}
\label{U^2}
\langle U(r)U(r)\rangle = \frac{e^2 d^2 \rho}{3} \int \frac{d^3 R}{R^4}=
\frac{4 \pi e^2 d^2 \rho}{3 r_0 } .
\end{equation}
 Here we assumed that $\langle d_i^{\alpha}d_j^{\beta} \rangle =d^2/3$ 
 $ \delta_{il} \delta_
{\alpha \beta}$ where $\alpha$ and $\beta$ are space indices,
and  the dipoles are distributed in space with a constant
density $\rho$. 
We had to introduce a cut-off parameter $r_0$ since the integral
diverges at small $R$. This parameter is, in fact,  the geometrical
size of the dipole. Indeed, inside the radius  $r_0$ the electric
field cannot be described by the dipole formula and the real potential $U(r)$
does not contain
the singularity $1/r^2$ which leads to the divergence of the integral.
Our cut-off corresponds to a zero field inside the sphere of radius $r_0$
\cite{comment}.

The correlator of the potentials at the points $\bf{r_1}$ and $\bf{r_2}$
 is equal to
\begin{equation}
\label{UU}
\langle U(\bf{r_1}) U(\bf{r_2})\rangle
= e^2\sum_{i,j} <{\bf d_i \nabla} \frac{1}{|\bf{r_1-R_i}|}
 {\bf d_j \nabla} \frac{1}{|\bf{r_2-R_j}|}>= \frac{e^2 d^2}{3} \sum_{j}
\left( {\bf \nabla} \frac{1}{|\bf{r_1-R_j}|}\right)
\left( {\bf \nabla} \frac{1}{|\bf{r_2-R_j}|}\right) .
\end{equation}

   Assume that the dipoles are distributed in space with a constant
density $\rho$. Then we have
\begin{equation}
\label{UUrho}
\langle U(\bf{r_1}) U(\bf{r_2})\rangle
= \frac{e^2 d^2 \rho}{3} \int d^3 R
\left ({\bf \nabla} \frac{1}{|\bf{r_1-R}|} \right )
\left ({\bf \nabla} \frac{1}{|\bf{r_2-R}|}\right )=
\frac{4 \pi e^2 d^2 \rho}{3|\bf{r_1-r_2}| } . 
\end{equation}
The integral here is convergent and can be easily calculated for $r_0=0$ using
integration by parts. It is interesting that the result is the same (does not
depend on $r_0$)
for any $r_0 < |\bf{r_1-r_2}|/2$. Indeed, this problem is mathematically
 equivalent
to the calculation of the electrostatic energy of two charged spheres
of the radius $r_0$ ( the interaction energy is proportional to the cross
term $2 \bf{E_1 E_2}/8 \pi$ in the energy density of electric field
 $\bf{E}^2/8 \pi$; $\bf{E_{1,2}} =
e\bf{\nabla}\left(1/|\bf{r_{1,2}-R}|\right)$). The answer is known: the
 interaction
energy is equal to $e^2/|\bf{r_1-r_2}|$  if  $2 r_0 < |\bf{r_1-r_2}| $.

Thus , we obtained the following result for the normalized correlator
\begin{equation}
\label{xi}
 \xi(k) \equiv \frac{\langle U(\bf{r})U(\bf{r+k})\rangle }
{\langle U^2\rangle }=\frac{r_0}{k} .
\end{equation}
We see that the correlations in the dipole random potential
decay inversely proportional to the distance.

  In the Refs. \cite{Grin1988,Luck1989,Izrailev1998} the following
expression for the inverse localization length for the 1D discrete 
Shrodinger equation
\begin{equation}
\label{psi}
\psi_{n+1}+\psi_{n-1}=(E+\epsilon_n)\psi_n
\end{equation}
has been obtained: 
\begin{equation}
\label{l}
l^{-1}=\frac{\epsilon_0^2\phi(\mu)}{8 \sin^2 \mu};
\end{equation}
\begin{equation}
\label{phi}
\phi(\mu)=1+ 2 \sum_{k=1}^{\infty} \xi(k) \cos(2 \mu k)
\end{equation}
Here the eigenenergy is $E=2 \cos\mu$, $\epsilon_n=U(r_n)$, $\epsilon_0^2=
\langle U^2\rangle $. This equation has been derived in the approximation
 $\epsilon_n \ll 1$. 
Now we can  substitute the correlator $\xi(k)=r_0/k$
 into this equation. The result is
\begin{equation}
\label{phidip}
\phi(\mu)=1- 2 r_0 \ln |2\sin\mu| . 
\end{equation}
In this equation $r_0$ is measured in units of the lattice constant.
 The minimum of $\phi$ is given by $\phi_{min} \simeq 1- 1.4 r_0$.
The delocalization corresponds to $\phi=0$ ($r_0=0.72$). This condition
seems to be impossible to satisfy. Indeed, the equation  $\xi(k)=r_0/k$
 is valid
for $r_0 < 0.5$. The realistic value of $r_0$ is  smaller than this limit.
A typical dipole size in  molecules
is about one Bohr radius while the lattice constant is about five times
 larger. This gives an estimate $r_0 \sim 0.2$. Also, any short-range
fluctuations increase $\langle U^2\rangle $
 and reduce the normalized long-range
correlator $\xi(k)$, see eq. (\ref{xi}). 

   However,  the correlations  change the localization length
significantly. It is well known that in the 2D case the localization length
is very sensitive to parameters of the problem. It would be
a natural guess that in the 2D case the correlations (\ref{xi}) due to
 the long-
range character of the dipole field are 
 more important
than in the 1D case, and that they may lead to delocalization. 

{\bf Acknowledgments.}
The author  acknowledges
the support from Australian Research Council. He is grateful  to F. Izrailev
and A. Krokhin for  discussions
and to  V.Zelevinsky for valuable comments 
and hospitality 
during the stay in MSU Cyclotron laboratory when this work was done.

\end{document}